\documentstyle[prl,aps]{revtex}
\begin{document}

\draft

\title{Entanglement and thermodynamical analogies}

\author{Pawe\l{} Horodecki \cite{mailp}}

\address{Faculty of Applied Physics and Mathematics\\
Technical University of Gda\'nsk, 80--952 Gda\'nsk, Poland}
\author{Ryszard Horodecki \cite{mailr} and Micha\l{} Horodecki \cite{mailm}}

\address{Institute of Theoretical Physics and Astrophysics\\
University of Gda\'nsk, 80--952 Gda\'nsk, Poland}

\maketitle

\begin{abstract}
We provide some new properties of entanglement of formation.
In particular, we obtain an additive lower bound
for entanglement of formation.
Subsequently we develop the concept of local orthogonality of
ensembles which leads to the mixed states with
distillable entanglement equal to entanglement of formation.
Then we consider thermodynamical analogies within the entanglement processing
domain. Especially, we exploit analogy entanglement -- energy. In this scheme
the total entanglement  i.e. the amount of singlet pairs needed for local
preparation of a state corresponds to internal energy while the free
entanglement defined as the number of pairs which can be recovered from the
state (distillable entanglement) is the  counterpart of free energy.
In particular, it leads us to the question about ``temperature'' of
entanglement. We also propose a scheme of the search of representative state
for given entanglement which can be viewed  as an analogue of  the Jaynes
maximum entropy principle.
\end{abstract}

\section{Introduction}
Quantum entanglement discovered
by Einstein, Podolsky, Rosen (EPR) \cite{EPR} and
Schr\"odinger \cite{Sch}, is one of the most
interesting quantum phenomena leading to profoundly
nonclassical effects revealed by contemporary physics
\cite{Ekert,dense,Tel1,comput,Cleve}.
The entanglement of pure quantum
state of composite system is defined as impossibility of factorisation
of the state. The simplest and, at the same time, the paradigmatic example of
such a state is the  singlet state of two spin-${1 \over 2}$ particles labelled
by $A$ and $B$
\begin{equation}
\Psi_{-}={1\over \sqrt2} (|\uparrow\rangle_{A}|\downarrow\rangle_{B}-
|\downarrow\rangle_{A}|\uparrow\rangle_{B}),
\label{EPR}
\end{equation}
or any state which can be obtained from the above one by means of
product unitary transformation.
The properties of the states of this kind are responsible for profoundly
nonclassical phenomena like quantum cryptography via Bell inequalities
\cite{Ekert}, quantum dense coding \cite{dense},
quantum teleportation \cite{Tel1}, quantum
computation \cite{comput} and the reduction of communication complexity
\cite{Cleve}. In real world we
usually meet physical systems which interact with environment getting then
entangled with it. This process, changing the state of our system from pure
to mixed one, described by  density matrix, decreases the internal
entanglement of the system sometimes even destroying it completely. Thus one
usually faces the following undesired process
\begin{equation}
| \Psi_{-} \rangle \langle \Psi_{-} |
\ \ \longrightarrow\ \ \varrho_{AB}.
\label{szum}
\end{equation}
Some time ago it has been shown \cite{Werner,Pop1,Pop2} that there are cases
when the system in a mixed state $\varrho_{AB}$ still possesses some residual
entanglement and physical effects
connected with this fact have been discovered
\cite{Pop1,Pop2,Gisin}. The density matrices of systems possessing residual
entanglement are called inseparable. Mathematically, the state
$\varrho_{AB}$ defined on the space
${\cal H}= {\cal H}_{A} \otimes {\cal H}_{B}$
is called inseparable (separable) if it cannot (can) be represented as a
convex combination of product states
\begin{equation}
\varrho_{AB}=\sum_{i=1}^Kp_i\varrho_{A}^{i}\otimes \tilde{\varrho}_{B}^{i}
\label{1}
\end{equation}
where $\varrho^i_{A}$ and $\tilde{\varrho}^i_{B}$ are
states of the subsystems and $\sum_{i=1}^K p_i=1$.
If the dimensions of the spaces ${\cal H}_{A}$, ${\cal H}_{B}$
are finite then the states $\varrho_i$ and $\tilde \varrho_i$
can be taken to be pure and one can consider only
the case $K\leq(\dim {\cal H}_{A} \dim {\cal H}_{B})^{2}$
\cite{tran,VP}.

Let us now consider  the paradigmatic situation of two observers Alice and Bob
being in two distant laboratories. There is a source
of pairs of particles between two laboratories
which sends one member of any pair to each of them.
Alice and Bob are allowed to perform any quantum operations on particles
in their laboratories and communicate with each other via some
classical channel (say telephone).
Usually they are also allowed to discard some particles.
We shall refer to  all those operations as to LQCC ones.
As Alice and Bob can only interact classically,
then some operations are certainly unavailable for them. For example if
they share a pair of the particles which are unentangled, it is
impossible to entangle those particles with each other.
Now the basic task is to find the best Alice and Bob can do under the
above restrictions to reverse somehow the process (\ref{szum}).
This leads to the recent
idea of distillation (or purification) of noisy entanglement
via local quantum operations and classical communication \cite{pur,conc}.
In this context,
there have been recently some attempts to build an appealing analogy between
the domain of entanglement processing and thermodynamics
\cite{Pop,Holl,VP,minent}. In this paper we develop the thermodynamics-like
approach to entanglement. Especially, we exploit the analogy entanglement --
energy. The paper is organized as follows. In sec. I and II we describe
entanglement measures and the process of distillation of entanglement.
In sec. III we show that entanglement of formation may not change under
the  highly irreversible twirling operation. We also show that there exist
mixed states for which the process of local preparing from singlet pairs is
reversible i.e. the number of the used pairs can be completely recovered by
means of quantum local operations and  classical communication. This involves
the notion of {\it local orthogonality}. In sec. IV we describe the analogy
entanglement -- energy. Then we discuss in this context  the known
results on entanglement processing. In sec. V we propose a scheme of
obtaining the {\it representative state} for given value of entanglement.
The scheme can be viewed as some analogue of the Jaynes principle of
maximum entropy.

\section{Entanglement measures}
In this section we briefly review some of the recently introduced
entanglement measures \cite{Bennett,Knight,VP}. Namely there  was a natural
problem how to quantify the entanglement of any quantum state
$\varrho_{AB}$ of composite system. Referring to
the paradigmatic situation of two distant laboratories, there arise
natural postulates which must be satisfied by  any measure of entanglement $E$
\cite{Knight,VP,Pop}

(i)  $E(\varrho_{AB})\geq 0$ and $E(\varrho_{AB})=0$ if $\varrho_{AB}$
is separable,

(ii) $E(\varrho_{AB})$ is invariant under the product
unitary operations $U_{A}\otimes U_{B}$,

(iii) $E(\varrho_{AB})$ cannot be increased by any LQCC operation.

The last axiom should be understood in the
stronger averaged entanglement sense
i.e. that if we have $ \varrho_{AB}^{i}$
being outputs of some LQCC operation of required property
with probabilities $p_{i}$ then
$\sum_{i} p_{i}E(\varrho_{AB}^{i})\leq E(\varrho_{AB})$. We assume here
that the considered operation is trace preserving, i.e. it does not
involve postselection. In fact, if Alice and Bob selected the
subensemble of  pairs corresponding some given outcome, then
the density matrix of subensemble could be more entangled than
the initial state. However, the total average entanglement cannot exceed
the initial one.  If one adds the condition that

(iv) entanglement of $n$ singlet pairs (\ref{EPR})
is equal to $n$, then one gets that the measure should satisfy \cite{Pop}
\begin{equation}
E(|\Psi_{AB}\rangle \langle \Psi_{AB}|)=
S(Tr_{A}(|\Psi_{AB}\rangle \langle \Psi_{AB}|), \quad
\ \rm where \em \
S(\sigma)=-{\rm Tr} \sigma\log\sigma \rm \ is \ the \ von\  Neumann \ entropy
\label{addpur}
\end{equation}
being then {\it additive} on pure states (in this paper we use base-2
logarithms).
In general, if we label the Alice and Bob particles by $A_{i}$,
$B_{i}$ then the additivity of E is defined as:
\begin{equation}
E(\varrho_{A_{1}B_{1}} \otimes \varrho_{A_{2}B_{2}} \otimes ...\otimes
\varrho_{A_{n}B_{n}})=
E(\varrho_{A_{1}B_{1}}) + E(\varrho_{A_{2}B_{2}}) + ...+
E(\varrho_{A_{n}B_{n}})
\end{equation}
There are known two measures satisfying all four  conditions (i-vi)
recalled above. It is not known however whether
they are additive. However the  condition (iv) automatically
guarantees that they are additive on pure states.
First one is the entanglement of formation $E_{f}$
\cite{Bennett} which
must be first defined on
the quantum ensemble ${\cal E}=\{ p_{i}, |\Psi_{i} \rangle \}$ as
\begin{equation}
E_{f}({\cal E})=\sum_{i}p_{i}S(Tr_{A}(|\Psi_{AB}\rangle \langle \Psi_{AB}|)).
\end{equation}
Than the entanglement of
formation of state $\varrho_{AB}$
is defined as
\begin{equation}
E_{f}(\varrho_{AB})=
\min_{\cal E} E_{f}({\cal E})
\end{equation}
where minimum is taken over all ensembles realising the state
$\varrho_{AB}$. The above function has some interesting properties,
in particular the analytical formula for $E_{f}$
for an arbitrary two spin-${1 \over 2}$ system
has been recently provided \cite{Woot}.
It is not known whether the above function is additive.
This is the main obstacle for fully consistent interpretation
for this quantity as the  asymptotic number of
singlets per output pair needed to build given
inseparable state. That was the  reason of introducing
the definition of internal entanglement \cite{my2}
which we shall call {\it total entanglement}
contained in state $\varrho $. It is given in the following

{\it Definition .-  The {\it total entanglement} contained in state $\varrho$
is defined as
\begin{equation}
E_{tot}(\varrho)=\lim_{n \rightarrow \infty} { E_{f}(\varrho^{\otimes n}) \over n}.
\label{toten}
\end{equation}}

This measure has the interpretation of the average number of singlet
pairs (per output pair) needed to produce pairs in state $\varrho$.
There is an open question whether the total entanglement is
equal to the entanglement of formation. It is certainly  {\it not}
greater than the latter and is additive on tensor product of the same states
i.e. $E_{tot}(\varrho\otimes\varrho)=2E_{tot}(\varrho)$. If the entanglement
of formation also satisfies this condition then the two measures
are identical i.e. $E_{tot}=E_f$.
As the problem of additivity of $E_{f}$ is open,
there is the question whether it is possible that $E_{f}\neq 0 $
while $E_{tot}=0$. Below we provide simple {\it additive }
lower bound for the entanglement of formation
which excludes this curiosity for
some states.

{\it Proposition .- For any state $\varrho_{AB}$
the entanglement of formation is bounded by:
\begin{equation}
E_{f}(\varrho_{AB})\geq S(Tr_{C}\varrho_{AB})-S(\varrho_{AB}),
\ {\rm with }\  C=A {\rm \  or \ } B
\label{ogran}
\end{equation}}
{\it Proof .-}
The proof of the above inequality is a simple
implication of the concavity property \cite{Wehrl}
of the function $(S(\cdot)-S(Tr_{A}(\cdot))$.
It is interesting to note that
the left hand side of the inequality (\ref{ogran})
defines two additive functions of $\varrho_{AB}$
\begin{equation}
G_C(\varrho_{AB})=S(Tr_{C}\varrho_{AB})-S(\varrho_{AB}), \ C=A {\rm \  or \ } B
\end{equation}

For $2 \times 2 $ Werner states
any of those functions, if
positive, is the amount of entanglement which
can be distilled from the states
by means of hashing method \cite{Bennett}.
It is not known whether this result can be
extended to higher dimensions as there is no counterpart
of hashing method there.
Note that exact positivity of the function $G(\varrho_{AB})$
is {\it sufficient} condition for inseparability of states.
It is important fact that,
as we announced before, additivity of $G(\varrho_{AB})$
prevent us from  the strange situation when  $E_{tot}=0 $ for all
inseparable states the inseparability of which results
from exact positivity of $G(\varrho_{AB})$.
In fact, for $G(\varrho_{AB})>0$
we have $E_{tot}(\varrho)=\lim_{n \rightarrow \infty} { E_{f}(\varrho_{AB}^{\otimes n})
\over n} \geq \lim_{n \rightarrow \infty} {G(\varrho_{AB}^{\otimes n}) \over n}=
\lim_{n \rightarrow \infty} {nG(\varrho_{AB}) \over n}=
G(\varrho_{AB})>0$.

The other measure of entanglement
is  {\it relative entropy of entanglement} \cite{Knight,VP}.
This measure is defined as
\begin{eqnarray}
E_{r}(\varrho)=min_{\sigma_{sep}}
S(\varrho|| \sigma), \quad { \rm \ with \ }\quad
S(\varrho||\sigma)=Tr(\varrho \log\varrho-\varrho \log\sigma_{sep})
\end{eqnarray}
with minimum taken over all separable states $\sigma_{sep}$.
There is inequality relation between two the measures \cite{VP}
\begin{equation}
E_{r}(\varrho) \leq E_{f}(\varrho).
\end{equation}
This measure has  good topological properties and important probabilistic
interpretation (see Ref. \cite{VP}).

\section{Distillation of noisy entanglement
and notion of distillable entanglement}

In this section we describe briefly the concept of distillation
of mixed state entanglement. Recall that Alice and Bob start with sharing
some amount, say $n$, of particles in a given initial state
and are allowed to perform a sequence of
any LQCC operations to obtain with
the, in general, less amount ($m_n$) of pairs
of particles in states  arbitrarily close to singlets.
The asymptotic average amount of the output pairs per input
pair is called
{\it distillable entanglement} \cite{pur,Bennett}.
Although the main idea is clear,
the precise definition of this notion is still missing.
An important attempt of
strictly mathematical formulation of the definition is contained in
Ref.\cite{Rains} by means of separable superoperators.
In this paper it has been proved that,
using trace preserving separable superoperators,
of constant output dimension one can distill from the Werner $2\times 2$
states only {\it strictly} less amount of entanglement
than the entanglement of formation $E_f$ of the states.
This important feature of {\it any} two local action
has been also derived by means of  $E_{r}$ measure \cite{VP}
at  the additional assumption that  the relative entropy of entanglement is
additive on product of identical states (at present we do not know whether
it is true). As the proof in Ref. \cite{Rains} does not consider the
Alice and Bob actions in full generality, the result is still waiting for
rigorous proof. Here we present an attempt the quantitative definition of
distillable entanglement \cite{pur,Bennett,Rains} enriched by the various
dimension output condition. Given $n$ pairs of particles, each  in state
$\varrho$, Alice and Bob are allowed to perform
arbitrary finite sequences LQCC operations on state
$\varrho^{\otimes n}$ with
$N$ different final outputs $ k=1, ..., \tilde{M^{n}} $ with
possibly different dimensions $d_{k}^{n}$.
Any sequence of LQCC operations can be written  as  some
separable superoperator \cite{Rains}
(but {\it not}  conversely \cite{NwE}) which can, in general,
produce different output systems acting as
\begin{equation}
\varrho^{\otimes n} \rightarrow \varrho'_k= {1\over p_i}
\sum_{i=1}^{\tilde{N}^{n}}
A_i^{(k)}\otimes B_i^{(k)}\varrho^{\otimes
n}{A_i^{(k)}}^\dagger\otimes {B_i^{(k)}}^\dagger
\end{equation}
where
\begin{equation}
p_k={\rm Tr}\left(\sum_i A_i^{(k)}\otimes B_i^{(k)}\varrho^{\otimes n}
 {A_i^{(k)}}^\dagger\otimes {B_i^{(k)}}^\dagger\right)
\end{equation} is the probability of the outcome and the states $\varrho'_k$
are defined on different
Hilbert spaces $H_{k}^{n}$, dim $H_{k}^n=d_{k}^{n}$; here one also requires that
$ A_i^{(k)}, B_i^{(k)}: H^{\otimes n} \rightarrow H_{k}^{n} $ and
$\sum_{i,k} p_iA_i^{(k)}A_i^{(k)\dagger} \otimes B_i^{(k)} B_i^{(k)\dagger}=I$.

Now the entanglement which can be distilled by means of such a given
protocol  $\cal P$ is defined
as
\[
D_{{\cal P}}= \lim_{n\rightarrow \infty} {\overline{m}_n\over n}
\]
with $\overline{m}_n=\sum_kp_k \log d_k^n$. Here
we demand that the input states $\varrho'_k$ tend to
singlet states $\Psi^+_d={1\over\sqrt d}\sum_{i=1}^{d}|i\rangle\otimes|i\rangle$
on spaces $H^n_k$ for high $n$. Quantitatively, the closeness between
the states and $\Psi^+$s is measured by fidelity
$F=\langle \Psi^+_{d^n_k}|\varrho|\Psi^+_{d^n_k}\rangle$ which should tend to 1
for high $n$. The latter condition is called {\it high fidelity condition}.
The condition must be stated in such a way, that the input states
can be  directly used  for  different  purposes like e.g. teleportation.
It appears that it is not easy to provide proper form of
condition \cite{Rainpriv} in general case. Of course, for constant output
dimension (i.e. where $d^n_k=d_n$ for each $k$) the condition can be stated
unambiguously \cite{Bennett,Rains} as
\begin{equation}
\sum_kp_k F(\varrho_k')\rightarrow 1.
\end{equation}
Also, there is no problem with all the protocols existing so far, as the latter
always produce some number of two qubit pairs, so that the output dimensions
are powers of $2\times2$. Indeed, suppose that given the outcome $k$, we obtain
$m^n_k$ two-qubit pairs in joint state $\varrho'_k$.
Let $\varrho^k_l$ denote the state of $l$th pair. The condition can
be of the form
\[
\lim_{n\rightarrow\infty}\inf_{l,k} F(\varrho^k_l)\rightarrow 1
\]
Thus we simply demand that the state of each obtained pair has to
tend to singlet state for high $n$. Now, having defined (in somewhat
incomplete way) the amount of distilled entanglement with respect to a given
protocol
we can define \cite{Bennett} the distillable entanglement of the state
$\varrho$ by maximizing  $D_{\cal P}$ over all possible protocols $\cal P$.
\[
D(\varrho)=\max_{\cal P} D_{\cal P}
\]
There are some results on $D$ which are ``definition independent''
\cite{pur,conc,Bennett,my1}.
First, trivially we have $E_{tot}\geq D$ as we certainly cannot
obtain greater number of singlets than the one necessary to produce the
state. Otherwise we would be able to create singlets by means of LQCC.
The inequality immediately implies that also $E_f\geq D$.
Much more nontrivial result is that for pure
states  \cite{conc} $D(\Psi)=E_{f}(\Psi)=E_{r}(\Psi)=E_{tot}(\Psi)$.
It is also known \cite{my1} that for any state from $2 \times 2$ case
$D\neq 0$ iff $E_{f}\neq 0$ ($E_{r}\neq 0)$. As we have mentioned above there
have been provided \cite{Rains,VP} quite strong arguments supporting the
statement  that for $2 \times 2$ Werner states
\begin{equation}
 D < E_{f}.
\label{irr}
\end{equation}
Quite recently, additional surprising information has been
provided \cite{my2}. Namely, for the systems $N\times M\geq 8$ there exist
states for which $D=0$ while  still $ E_{f} > 0$.
The result has been achieved by proving that any distillable state must
violate Peres criterion of positivity of partial transposition \cite{Peres}
(see also \cite{sep})
and recalling that there are states
which  satisfy the criterion being still inseparable,
having then nonzero $E_{f},E_{r}$ \cite{tran}.
This result indicates the possibility of dramatic {\it qualitative}
irreversibility of the process production of mixed entangled states.
The entanglement needed to produce such states become completely bound, so
that no amount of it can be recovered by means of LQCC.
On the other hand, the inequality (\ref{irr}) represents the {\it
quantitative} irreversibility expected for some mixed states.
Since we know that for pure states the process of production
of pure not maximal entangled states is reversible
($D=E_{f}$) then the natural question is whether there are mixed states
for which still equality $D=E_{f}$ holds.
We shall discuss this question in detail in the next section.
Remarkably, the above  irreversibilities, which are
physically intuitive, could be regarded as
fully exact ones only if we knew that indeed $E_{f}=E_{tot}$.
We do not know it yet, but from the definition of $E_{tot}$
it is easily to show that this quantity is nonzero
whenever $D$ is nonzero, and is equal to $D$ if $D=E_{f}$.
In table I we collect the results concerning relation between
$D$, $E_{tot}$ and $E_f$.

\section{Local orthogonality concept
and mixed states with $D=E_{f}$}
\subsection{Illustrative example}
As we have mentioned in previous section
it has been argued recently \cite{Rains,VP} that distillable
entanglement $D$ is strictly less than the entanglement of formation $E$
for $ 2 \times 2 $ Werner states. The latter are of the form
\begin{equation}
\varrho_{W}(F)\equiv
F|\Psi_{+} \rangle \langle \Psi_{+}|+
\frac{1-F}{3}|\Psi_{-} \rangle \langle \Psi_{-}|+
\frac{1-F}{3}|\Phi_{+} \rangle \langle \Phi_{+}|+
\frac{1-F}{3}|\Phi_{-} \rangle \langle \Phi_{-}|.
\label{Wern}
\end{equation}
Here we use the usual Bell basis \cite{Mann}
\begin{eqnarray}
\Psi_{\pm}=\frac{1}{\sqrt{2}}(|00\rangle\pm |11\rangle \nonumber \\
\Phi_{\pm}=\frac{1}{\sqrt{2}}(|01\rangle\pm |10\rangle.
\label{Bell}
\end{eqnarray}
Below we provide some quite interesting property of the function $E_{f}$
which is compatible with the mentioned results. Consider any pure
state of the form
\begin{equation}
|\Psi \rangle= a|00\rangle +b |11\rangle , \ a,b >0, \ a^2+b^2=1.
\label{Psi}
\end{equation}
By definition the entanglement of formation of  (\ref{Psi}) amounts to
\begin{equation}
E_f(\Psi)=H(a^2), \end{equation}
where  $H(x)= - x\log x - (1-x) \log (1-x)$
is the binary entropy function.
Let us subject the above state to the following ``twirling'' operation
i.e. the random bilateral operation \footnote{
We use here second unitary transformation conjugated relative to
the first one for the sake of the further analogy in higher dimensions.
For $2\times 2$ case the original
twirling operation using the same unitary operations on the first and
the second subsystem can be also utilized.}
\cite{Werner,Bennett,pur}
\begin{equation}
{\cal T}(\cdot) \equiv \int U \otimes U^{*} ( \cdot ) U \otimes U^{*} dU
\end{equation}
where the integral is performed with respect to the probability
measure proportional to the Haar measure.
In result  we obtain a  Werner state
\begin{eqnarray}
{\cal T}(|\Psi \rangle \langle \Psi|)=\varrho_{W}(F) \quad {\rm with} \quad
F={(a+b)^2 \over 2}
\end{eqnarray}
It is elementary to see that
$ \frac{1}{2} + \sqrt{F(1-F)}=
\frac{1}{2}(1 + \sqrt{1-4a^2b^2}) = a^2$.
Now, since
$E(\varrho_{W}(F))=
H(\frac{1}{2} + \sqrt{F(1-F)})$ \cite{Bennett} we see that
\begin{eqnarray}
E({\cal T}(|\Psi \rangle \langle \Psi|))=
E(\varrho_{W}(F))=E(|\Psi \rangle \langle \Psi|).
\end{eqnarray}

Thus we have the curious situation than any pure
entangled state $|\Psi \rangle \langle \Psi|$ has {\it the
same} entanglement of formation as the highly randomized
state ${\cal T}(|\Psi \rangle \langle \Psi|)=\varrho_{W}(F)$.
Recall that for any pure state $|\Psi \rangle$
one has $D(\Psi)=E(\Psi)$.
Thus one can turn all the entanglement contained in $\Psi$
into the pure singlet form.
But for the Werner states (\ref{Wern})
it has been argued \cite{Rains,VP}
that  $D(\varrho_{W}(F))\leq \log2 - H(F) < E_f(\varrho_{W}(F))$
if only $F<1$. Thus the twirling operation in $ 2\times 2 $
case is probably an example of the operation preserving $E_f$
but significantly decreasing $D$. In fact, using the Schmidt decomposition
one can immediately prove the following

{\it Proposition .- For any $2\times 2$ pure state
there exists basis (given by Schmidt decomposition)
such that the operation ${\cal T}$ defined
in this basis preserves entanglement of formation. }

For $N \times N $, $N>2$ this proposition is {\it not} true.
To provide the counterexample let us consider
$N \times N $ system in the state $\Psi_{M}=\frac{1}{N}\sum_{i=1}^{M}|ii\rangle$,
$1<M<N$ having $E(\Psi_{M})=\log M $.
After the operation ${\cal T}$ one obtains the state \cite{MP}
\begin{equation}
\sigma(F)\equiv {N^{2} \over N^2-1}\left((1-F)
{I\over N^2}+(F-{1\over N^2})P_+\right),\ \ F=\frac{M}{N}.
\end{equation}
with $F\leq \frac{M}{N}$, and maximally entangled state
$P_+=\frac{1}{N}\sum_{i=1}^{N}|i\rangle\otimes |i\rangle$.
The above state can be represented as
$\sigma(F)=\frac{N(1-F)}{N-1}\sigma(\frac{1}{N})+
\frac{NF-1}{N-1}P_+$. As the state $\sigma(\frac{1}{N})$
is separable \cite{MP} we get immediately that
\begin{equation}
E(\sigma(F))\leq \frac{NF-1}{N-1} \log N =
\frac{M-1}{N-1} \log N < \log M, \ \ {\mbox for} \ \ 1< M < N.
\end{equation}
Thus in higher dimensions there exist the pure entangled
states for which the random bilateral unitary
transformation always decreases the entanglement of formation.
Thus the proposition provides the {\it unique}
feature of two-qubit entanglement.
Note in this context
that the results concerning distillation of entanglement
stress that, in a precise sense,
any distillable entanglement is a mixed two-qubit entanglement
(see \cite{my2} for details).

\subsection{Local orthogonality}
The expected and partially proved fact
that the operation ${\cal T}$  decrease D
may be supported by the
intuition that it should not be possible
to distill the full entanglement content from any of
pure states $U\otimes U^{*}\Psi$ included in the mixture
${\cal T}(|\Psi \rangle \langle \Psi|)$
as they can not be perfectly distinguished
by local operations and classical comunication.
Recently this fact has been used in
the proof of imposibility of local cloning
of Bell states (\ref{Bell}) \cite{Tal}.
The essential idea of the latter was the
observation that any two orthogonal entangled states can be
cloned locally only if some of their reduced density matrices
are orthogonal. We shall call this property
{\it local orthogonality} proposing the general definition

{\it Definition.-
Consider two
states $\psi$, $\phi$ of composite multiparticle composite quantum system
defined on Hilbert space ${\cal H}=\mathop{\otimes}  \limits_{l=1}^{m}
{\cal H}_l$.
We say that the two states are  {\bf k-locally orthogonal}, if there exist
some $k$ subsystems
(say they are labelled by $\{i_1,\ldots, i_k\}$ and are described by
Hilbert spaces ${\cal H}_{i_1},\ldots, {\cal H}_{i_k}$) such that
the corresponding reductions of the states
$\psi$, $\phi$ are orthogonal, i.e.
\begin{equation}
Tr(\varrho^{\psi}_{l} \varrho^{\phi}_{l})=0, \quad
l=i_{1}, ...,i_{k}.
\end{equation}
If the numbers of the systems $\{i_1,\ldots, i_k\}$ are known
then the states $\psi$, $\phi$
are called {\bf locally orthogonal on the subsystems
$\boldmath{i_1,\ldots, i_k}$}.
The ensemble of pure
states $\{ \psi_{i} \}_{i=1}^{K} \}$ is called {\bf locally orthogonal}
if its elements can be ordered in the sequence
$\{ \psi_{i_{1}},\psi_{i_{2}}, ...,
\psi_{i_{K}}\}$ such that for any $1\leq m \leq N$
the state
$\psi_{i_{m}}$ is 1-locally orthogonal
on the same subsystem
to all following states $\psi_{i_{n}}, n > m $.}
\vskip1mm

{\it Remark.-} The notion of local orthogonality is not equivalent to the
notion of  local distinguishability. The states in a given  ensemble can
be locally distinguishable, but it may be the case, that to
distinguish them one must destroy them
(see in this context \cite{NwE}). The locally orthogonal states
are distinguishable without destroying them.
\vskip1mm

Note that any two component quantum system
allowing for existence two {\it entangled} 1-locally orthogonal
pure states must be of the form $M \times N$, $\max (M,N)\geq 4$
i.e. one of subsystems must represent at least the spin
$\frac{3}{2}$.
We have the following simple

{\it Property .- Consider the state
$\varrho= \sum_{i}p_{i}|\Phi_{i} \rangle \langle \Phi_{i}|$ of
the quantum system composed from two subsystems
defined by locally orthogonal ensemble $\{ \Phi_{i} \}$.
Then we have

(i) $E_{f}(\varrho)=\sum_{i}p_iE_{f}(|\Phi_{i}\rangle \langle \Phi_i|)$

(ii) $D(\varrho)=E_{f}(\varrho)=E_{tot}(\varrho)$.}

{\it Proof .-} Consider the system in state $\varrho$
in  paradigmatic situation of two distant laboratories and
let us treat the state $\varrho$ as a random mixture of $\Phi_{i}$s.
Then, using appropriate local
measurements (following from local orthogonality
property of the ensemble)
and classical communication
Alice and Bob can determine which of the
pure state they share.
Then they can use the large blocks procedure \cite{conc}
to convert {\it all} the entanglement of this state
into the singlet form D. Thus
we have $D=\sum_{i}p_iE_{f}(|\Phi_{i}\rangle\langle\Phi_i|)$. As, by
definition,  $D(\varrho)\leq E_{f}(\varrho) \leq \sum_{i}E_{f}(|\Phi_{i}\rangle\langle
\Phi_i|)$ we obtain both numerical value of
$E_{f}$ and its equality to $D$. As the latter is, again by definition, additive
on products of the same states and as $E_{tot}\leq E_{f}$
we get easily from the first equality of (ii) the second one.

The simple example of $5 \times 5 $ locally orthogonal ensemble
ordered as required in the definition is:
$\{ |00\rangle + |11\rangle, |21 \rangle +|32\rangle,
|30\rangle + |43 \rangle $.
There remains an interesting question of maximal support
of the locally orthogonal ensembles for fixed composite quantum
system. In the above context we propose the two following
conjectures:

{\it Conjecture 1.- } The locally orthogonal ensembles are the
only ones which pure entangled components $\Psi_{i}$
can be distinguished by local operations and classical
communication without destroying them.

{\it Conjecture 2.- } The only mixed states $\varrho$
with the property $E_{tot}(\varrho)=D(\varrho)$
are those which are defined by means of locally
orthogonal ensembles.

Finally, note that the result of providing mixtures with $D=E_{tot}$ is
analogous to that of Braunstein, Mann and Revzen \cite{Mann}
who found  mixtures violating maximally the Bell inequalities. Their
mixtures were, in our language, locally orthogonal mixtures of singlet pairs.

\section{Thermodynamical analogies}
The first formal thermodynamical analogy was proposed
by Popescu and Rohrlich \cite{Pop}. The authors made an important observation
that any LQCC process which preserves entanglement
must be {\it reversible} which have been related
to Carnot cycle in usual thermodynamics. Such  an analogy was also considered
by Vedral and Plenio  \cite{VP} in the context of 
distillation of entanglement. In more general quantum information
context the thermodynamical approach was also
developed in Ref. \cite{Holl} where, in particular, the law  of conservation
of quantum information (entanglement) was considered. The analogy
entanglement-energy was then developed in Ref. \cite{my2} to interpret  the
results on distillation of entanglement. Below we propose some new elements
which, we believe, will contribute to understanding of thermodynamics-like
aspects of entanglement processing.

From previous discussion
we know that there are two important measures of
entanglement which are in a sense dual ones:

\begin{itemize}
\item Total entanglement $E_{tot}(\varrho)$ which represents
 the least number of shared singlets asymptotically required to prepare
 the state $\varrho $ by means of LQCC.
\item Distillable entanglement  $D(\varrho)$ - the greatest number of pure
singlets that can asymptotically be prepared from $\varrho$
by means of LQCC.
\end{itemize}
In short, $E_{tot}$ is the minimal number of singlets we need to produce a  state
while $D$ is the maximal number of singlets we can recover from the state.
Another important quantity of precise information theoretic sense is
von Neumann entropy $S$. Its physical sense in quantum information theory was
proved by Schumacher \cite{Schumacher} (see also Ref. \cite{JS}) and Barnum
{\it et al.} \cite{Barnum} in the context of compressing of
quantum information.
Basing on the above notions we shall now try to provide
infor\-ma\-tion-the\-ore\-tic
counterparts of such thermodynamical notions as internal energy $U$,
free energy $F$ and entropy $S_{th}$.
We propose the following scheme
for the ``energetic'' quantities \cite{my2} $E_{tot}$, $D$.
\vskip2mm
\centerline{(free entanglement)\ \ \   {\Large $D=E_{free} \leftrightarrow F$}\ \ \ (free energy)}
\vskip2mm
\centerline{(total entanglement) \ \ {\Large $E_{\scriptscriptstyle \rm tot}
\leftrightarrow U$}\ \  (internal energy)}
\vskip2mm
\centerline{(bound entanglement)  \ {\Large $E-D\equiv E_{\scriptscriptstyle\rm bound}
\leftrightarrow
TS_{th}$} \ (bound energy)}
\vskip2mm
where $T$ is temperature.
Now the last element of the above scheme lead us to the question what about
the entanglement counterparts of thermodynamical temperature and entropy?
Following analogy we obtain a formula
\begin{equation}
E_{tot}=E_{free}-T_e S_e
\label{free}
\end{equation}
where $T_e$ and $S_e$ are unknown counterparts of $T$ and $S_{th}$.

If one tries to recognize the von Neumann entropy $S$ as the counterpart
of thermodynamical entropy then finding the interpretation
of temperature of entanglement $T_{e}$  would be the main test of this choice.
It can be partially verified that the choice might be reasonable.
First, note that in thermodynamics one has $F_{th}\leq U_{th}$. In our
case the analogous inequality holds, as trivially, $D$ cannot exceed $E_{tot}$.
Now, recall that in usual thermodynamics $ F=U-TS_{th}$.
According to our proposal $ E_{free}=E_{tot}-T_{e}S $
even if we do not know what the temperature $T_{e}$ means.
Suppose now, that for pure states $T_{e}$ is finite. Then putting
$S=0$ we obtain that $D=E_{tot}$, i.e. that for pure states, the
distillable entanglement is equal to the entanglement of formation, which is
true indeed \cite{conc}.
Note, that for mixed states, the formula (\ref{free})
can serve as a definition of temperature $T_{e}$ for mixed states $\varrho$:
\begin{equation}
 T_{e}(\varrho_{mix})={E_{tot} - E_{free} \over S}.
\end{equation}
As for separable mixed state $\varrho$ we have $E=D=0$ we obtain that
the temperature vanishes in this case which can be extrapolated via expected
continuity property to cover the case of pure separable states.
So far the von Neumann entropy seems to be a good quantity
for our purposes. But, as we shall see in a moment, there are
some problems.
Note that for many mixed inseparable states, the best known protocols
of distillation provide a very little number of distilled singlets in
comparison with the entanglement of formation. For example for Werner states
the obtained yield of distillation procedure is about thousand times less than
the entanglement of formation. There are even more
stronger suggestions (recalled in the previous sections)
that for mixed $2 \times 2$ inseparable states $E_{free} < E_{tot}$
(provided that $E_{tot}=E_f$).
Now, in our language, this is equivalent to state
that for those states the temperature is nonzero.
However, as we have shown in  sec. IV
for composite systems of higher dimensions there
are many states with $E_f=E_{tot}=E_{free}$ and $S\neq 0 $.
This is a curiosity of the model and it would suggest that
the temperature of those states is zero which is not intuitive.
Thus perhaps the entropy $S_e$ should be
defined in other way. For this purpose one could exploit the
notion of local orthogonality. Namely,  the entanglement entropy $S_e$ should
quantify the irreversibility of the process of local preparing of the given
mixed state. Then it could be defined as some measure of local non-orthogonality
of some  {\it canonical} ensemble realizing the state. A natural candidate is
here the optimal ensemble in formula defining the entanglement of formation.
So defined entropy would vanish for locally orthogonal mixtures.
However, it is a very hard task to obtain a precise definition, and it is
an open question, whether that choice would produce the entanglement
temperature having a good physical interpretation.
Nevertheless, the energy type analogies about
$D$ as a counterpart of free energy and
$E_{tot}$ as the counterpart of internal energy seem to be plausible.

The remaining question is to define the entanglement (informational)
work. In general, it is natural to assume that
\begin{eqnarray}
\centerline{\large sending qubits $\leftrightarrow$ WORK} \nonumber
\end{eqnarray}
Indeed, suppose that the given system contains some amount of free entanglement.
Then one can use it to send the this precisely this amount of qubits by
means of teleportation preceded by distillation procedure. This is analogous to
the situation in thermodynamics, where the free energy can be used
to perform work. On the other hand, if the work $\Delta W$ is performed
over a system (without dissipation i.e. adiabatically) then the
total energy increases $U_2=U_1+\Delta W$. This corresponds to the fact that
to produce a state of total entanglement $E_{tot}$ Alice and Bob need to
exchange precisely this amount of qubits (they will send halves of singlets).
This is true if the channel between Alice and Bob is noiseless which
correspond to lack of dissipation of energy. If, instead, the channel is noisy,
then the number of qubits must be  larger: some amount of the sent entanglement
will be {\it spread} over the system and environment
\footnote{Note that entanglement can be spread deliberately by means of a kind
of ``depurification'' procedure \cite{Buzek}}. Then, despite the
total entanglement (total quantum information) of system plus
environment is conserved, the amount of {\it useful} entanglement
is much more less than the number of
exchanged qubits. We may say that there is the ``informational heat flow'':
the lost entanglement was not used to perform work, but rather
changed into uncontrolled form.
In the context of the possibility of existence of states with nonzero
bound entanglement the process of this flow can manifest itself not only by
destroying some amount of entanglement
but also in binding some part of remaining entanglement.
The above consideration is nothing else but a balance analogous to that
governed by the first law of thermodynamics.
As the latter is nothing else but
the principle of conservation energy, we obtain that the above balance
is implied by the conservation of quantum information
which can be viewed as the
analogue of the thermodynamical law  for quantum entanglement processing
\cite{Holl}.

The reader can ask why sending qubits and sending entangled qubits can both
be treated as work. Indeed there is a  clear qualitative difference between
them. However, it can be explain as follows: sending {\it known}
entangled qubits can be represented as  work {\it over} the composite system
i.e. as a  counterpart of mechanical work done over the gas,
as  it aims at  increasing of entanglement - ``energy'' of the system. The
latter can be subjected to  dissipation which causes the flow discussed
above. On the other hand, having the composite system
with given  entanglement $E_{free}$ one
can teleport some unentangled qubit through it.
This represent the {\it work done by the system} for us, resembling the gas
doing mechanical work. One can also send the unentangled qubits directly
without using entanglement and teleportation.
Then for noiseless channel we have a counterpart of
purely mechanical process (with no thermodynamical element).

\section{Searching for representative state at given entanglement}
\label{gibbs}
In this section we consider the problem of choosing for each established value
of entanglement, some representative, most probable state.
Such a choice can be of course performed only  up to local
unitary transformation. As a criterion we will use von Neumann entropy, so
that the scheme of obtaining the representative state will slightly resemble
the Jaynes scheme \cite{Jaynes} of producing Gibbs state. Recall that given
the Hamiltonian
H the Gibbs state $\varrho_G= (1/ {\rm Tr} e^{\beta H}) e^{\beta H}  $ for
a given  mean energy $E=\langle  H \rangle ={\rm Tr} \varrho H$ can be obtained
by maximizing von Neumann entropy over all states with mean energy $E$.
In our scheme, we will keep constant entanglement and maximize entropy (see
\cite{minent} in this context).
As it is total entanglement which we chose as the  counterpart of
energy, we should use this measure in our scheme. However, to be able
to perform any calculations we must have analytical formula for
entanglement. In this situation we will rather use entanglement of formation.
In the case of two-qubit states the analytic formula for the
latter is the following \cite{Woot}
\begin{equation}
E_f(\varrho)=H({1+\sqrt{1-C^2}\over 2}).
\end{equation}
Here
$C=\max\{0,\lambda_1-\lambda_2-\lambda_3-\lambda_4\}$
and $\lambda_i$s are the eigenvalues, in decreasing order, of the
Hermitian matrix $R=\sqrt{\sqrt\varrho\tilde\varrho\sqrt\varrho}$
with
\begin{equation}
\tilde\varrho=\sigma_y\otimes\sigma_y \varrho^* \sigma_y\otimes \sigma_y
\end{equation}
The star denotes complex conjugation of the matrix $\varrho$ in product basis.
Now, one could ask what state  is expected to have the greatest von Neumann
entropy of all states with given entanglement of formation.
Such a state should have a high degree of symmetry. The natural candidate
is the Werner state (\ref{Wern}) as it can be written in the following
very symmetric form
\begin{equation}
\varrho_W=\alpha |\Psi_{+}\rangle\langle\Psi_+| +(1-\alpha) {1\over 4} I\otimes I
\quad {\rm \ where \ } \quad {1\over3} \leq\alpha\leq1,
\end{equation}
Surprisingly, we will see that for some values of $E_f$ the
representative state
is certainly not the Werner state. Consider the following simple state
\begin{equation}
\varrho_p=p|\Psi_+\rangle\langle\Psi_+|+(1-p) |01\rangle
\langle 01|,
\end{equation}
where $\alpha={1\over3}(4F-1)$.
It can be checked that it is entangled for any $p>0$
\cite{Bennett,Peres,sep}. We obtain that
\begin{equation}
C(\varrho)=p,\quad C(\varrho_W)=2F-1
\end{equation}
Then the two states have the same entanglement of formation if
$p=2F-1$. This is compatible with the fact that for $p=F=1$ both of them
are equal to $|\Psi_+\rangle\langle\Psi_+|$ while for $p=0, F={1\over2}$
both of them are separable. Now, one can check that at least
for ${3\over4}<p<1$ the entropy of the state $\varrho_p$ is strictly greater
than the one the Werner state. Consequently, it is certainly not the Werner
state which maximizes entropy for those values of entanglement. Then
 one can conclude that  the problem of finding the representative
state for given entanglement may produce some highly nontrivial
result: it will provide an interesting, unknown  family of states.

\section{Concluding remarks}
In conclusion, we have provided new properties of 
entanglement of formation. In particular an additive lower bound for this 
quantity has been provided. The concept of local orthogonality have been 
developed leading to the family of mixed states with distillable 
entanglement equal to entanglement of formation.  Subsequently the notions 
of total and distillable entanglement have been considered as counterparts of 
thermodynamical notions of internal and free energy. The question of possible 
temperature of entanglement as well as the counterpart of thermodynamical 
entropy have been analysed. The process of sending quantum information has 
been considered as a counterpart of work and discussed in detail.  

One of the advantages of the proposed approach
is certainly the fact that it generates new interesting questions like
e.g. the problem of defining and interpretation of temperature of entanglement.
Another, fundamental  problem is the following: is there a link
between the analogy considered here and the recent
development concerning quantum information processing at incomplete data
\cite{minent,Jayncom}? In fact,  since the famous Jaynes papers we know that
the statistical thermodynamics which explains the phenomenological one can
be treated as a special kind of statistical inference at incomplete
experimental data. Then, it follows that the recent results on quantum
information processing at incomplete data (where one requires some
nonstandard schemes \cite{minent}) should also  be somehow connected with
the approach discussed in this paper. We think that since  analogy was
always a powerful tool in
physics, the above problems are worth of deeper investigation. We believe that
the present consideration will contribute to obtaining more clear picture
of the highly nonintuitive domain which is quantum information theory.

The authors would like to thank Eric Rains for helpful discussion
on the definition of distillable entanglement. They are also grateful to
Charles Bennett, Hans Briegel, Nicolas Cerf, Chris Fuchs and Tal Mor
for useful discussions. M. H. and P. H. kindly acknowledge the support
from Foundation for Polish Science.

\begin{table}
\caption[Coefficients]
{\label{wsp} Comparison of relations between $D$ and $E_{tot}$, $E_f$.}
\begin{center}
\begin{tabular}{ccc}
\hfill  
&$M \times N $ states \hfill &\\
Low dimension - $N\times M \leq 6$ && Higher dimension - $N\times M > 6$  \\
\hline
\hline  
\hfill& $\forall \varrho $ pure $D=E_{tot}=E_f$ &\hfill \\
\hline
$\forall \varrho$ mixed
$E_f>0 \Rightarrow  D>0$ & &$ \exists \varrho$ (mixed) with $E_f>0 $
and $ D=0 $ \\
\hline
Probably $ \forall \varrho$ mixed $D < E_{tot}$ if only $ E_{tot}>0 $
& & $\exists $ (locally orthogonal) $\varrho$ mixed with $D=E_{tot}=E_f\neq 0$
\\
\end{tabular}
\end{center}
\end{table}

\end{document}